\documentclass[prl,reprint]{revtex4-1}
\usepackage{graphicx}
\usepackage{color}
\usepackage{bm}
\usepackage{hyperref}

\begin{document}

\title{Spin-Polarization of Composite Fermions and Particle-Hole Symmetry Breaking}
\date{today}

\author{Yang Liu$^{1}$, S.\ Hasdemir$^{1}$, A.\ W\'ojs$^{2}$, J.K.\
  Jain$^{3}$, L.N.\ Pfeiffer$^{1}$, K.W.\ West$^{1}$, K.W.\
  Baldwin$^{1}$, M.\ Shayegan$^{1}$} \affiliation{Department of
  Electrical Engineering, Princeton University, Princeton, New Jersey
  08544} \affiliation{Institute of Physics, Wroc\l aw University of
  Technology, 50-370 Wroc\l aw, Poland} \affiliation{Department of
  Physics, 104 Davey Lab, Pennsylvania State University, University
  Park PA, 16802}

\date{\today}

\begin{abstract}
  We study the critical spin-polarization energy ($\alpha_{\rm C}$)
  above which fractional quantum Hall states in two-dimensional
  electron systems confined to symmetric GaAs quantum wells become
  fully spin-polarized.  We find a significant decrease of
  $\alpha_{\rm C}$ as we increase the well-width. In systems with
  comparable electron layer thickness, $\alpha_{\rm C}$ for fractional
  states near Landau level filling $\nu=3/2$ is about twice larger
  than those near $\nu=1/2$, suggesting a broken particle-hole
  symmetry. Theoretical calculations, which incorporate Landau level
  mixing through an effective three-body interaction, and finite layer
  thickness, capture certain qualitative features of the experimental
  results.
\end{abstract}


\maketitle

A hallmark of an interacting two-dimensional electron system (2DES)
subjected to a strong perpendicular magnetic field ($B$) are the
fractional quantum Hall states (FQHSs) which are observed
predominantly at odd-denominator Landau level (LL) fillings $\nu$
\cite{Tsui.PRL.1982, Jain.CF.2007}. They stem from the strong Coulomb
interaction energy between electrons, $V_{\rm C}=e^2/4\pi\epsilon
l_B$, and signal the formation of incompressible electron liquid
states with strong short-range correlations ($\epsilon$ is the
dielectric constant and $l_B=\sqrt{\hbar/eB}$ is the magnetic
length). The composite Fermion (CF) theory, in which two magnetic flux
quanta are attached to each electron, maps the interacting electrons
to a system of nearly non-interacting CFs \cite{Jain.CF.2007,
  Jain.PRL.1989, Halperin.PRB.1993}, and explains many properties of
FQHSs. For example, CFs have their own discrete,
magnetic-field-induced, orbital energy levels, the so-called
$\Lambda$-levels, whose filling factor is denoted by $\nu^{\rm
  CF}$. The integer quantum Hall states of CFs at $\nu^{\rm CF}$
manifest as FQHSs of electrons at filling $\nu=\nu^{\rm CF}/(2\nu^{\rm
  CF}+1)$ and $2-\nu^{\rm CF}/(2\nu^{\rm CF}+1)$.

The CFs also have a spin degree of freedom. In 2DESs confined to GaAs,
because of the small Land\'e $g$-factor ($g^*=-0.44$), the Zeeman
energy ($E_{\rm Z}$) is in fact comparable to the CF $\Lambda$-level
separation, which is equal to a small fraction of $V_{\rm C}$. Thus,
CFs might be partially spin-polarized when $E_{\rm Z} << V_{\rm C}$,
and become fully spin-polarized only if $E_{\rm Z}/V_{\rm C}$ is
larger than a $\nu$-dependent critical value $\alpha_{\rm C} \simeq
0.02$ which should be intrinsic to the 2DES and independent of the
sample quality \cite{Park.PRL.1998, Davenport.PRB.2012}. Moreover, in
an \textit{ideal} 2DES with zero layer-thickness and no LL mixing, the
FQHSs that have the same $\nu^{\rm CF}$, e.g. $\nu=2/3$ and 4/3
($\nu^{\rm CF}=-2$), are expected to have the same $\alpha_{\rm C}$
because of particle-hole symmetry $\nu\leftrightarrow 2-\nu$
\cite{Jain.CF.2007}. Experimentally, the spin-polarization of CFs has
been probed through measurements of transitions of FQHSs in both
transport and optical studies \cite{Eisenstein.PRL.1989,
  Engel.PRB.1992, Du.PRL.1995, Kukushkin.PRL.1999, Smet.PRL.2001,
  Hashimoto.PRL.2002, Kraus.PRL.2002,Tracy.PRL.2007,
  Groshaus.PRL.2007}. In these studies, $E_{\rm Z}/V_{\rm C}$ in
increased by either increasing the 2DES density or adding a parallel
magnetic field. However, no systematic measurements of $\alpha_{\rm
  C}$ on samples with controlled parameters, such as layer thickness,
are scarce \cite{Vanovsky.PRB.2013}.

Here we report measurements of $\alpha_{\rm C}$ for 2DESs confined to
GaAs quantum wells (QWs) and with carefully controlled,
\textit{symmetric} charge distributions \footnote{In our QW samples,
  $\alpha_{\rm C}$ shows a non-trivial dependence on the symmetry of
  the charge distribution, which needs to be studied carefully in the
  future.}. Moreover, in our experiments, to avoid charge distribution
distortions that occur when a strong parallel magnetic field is
introduced, we apply only perpendicular magnetic fields, and tune the
ratio $E_{\rm Z}/V_{\rm C}$ at a fixed $\nu$ via changing the 2DES
density. Our systematic measurements reveal that $\alpha_{\rm C}$
strongly depends on the charge distribution thickness. The thicker the
2DES is, the smaller $\alpha_{\rm C}$. We also find that, for a given
electron layer-thickness, normalized to $l_B$, $\alpha_{\rm C}$ is
much larger for the FQHSs around $\nu=3/2$ compared to their
particle-hole counterparts around $\nu=1/2$, implying that
particle-hole symmetry is broken. We present results of
state-of-the-art microscopic calculations based on CF wavefunctions,
The calculations include LL mixing through an effective three-body
interaction, which affects fractions $\nu$ and $2-\nu$ differently and
breaks the particle-hole symmetry. We find that, while the effect of
LL mixing is typically small on the energy of any given FQHS, it has a
substantial effect on the rather small energy {\em differences} that
determine the $\alpha_{\rm C}$.

\begin{figure}[htbp]
\includegraphics[width=.4\textwidth]{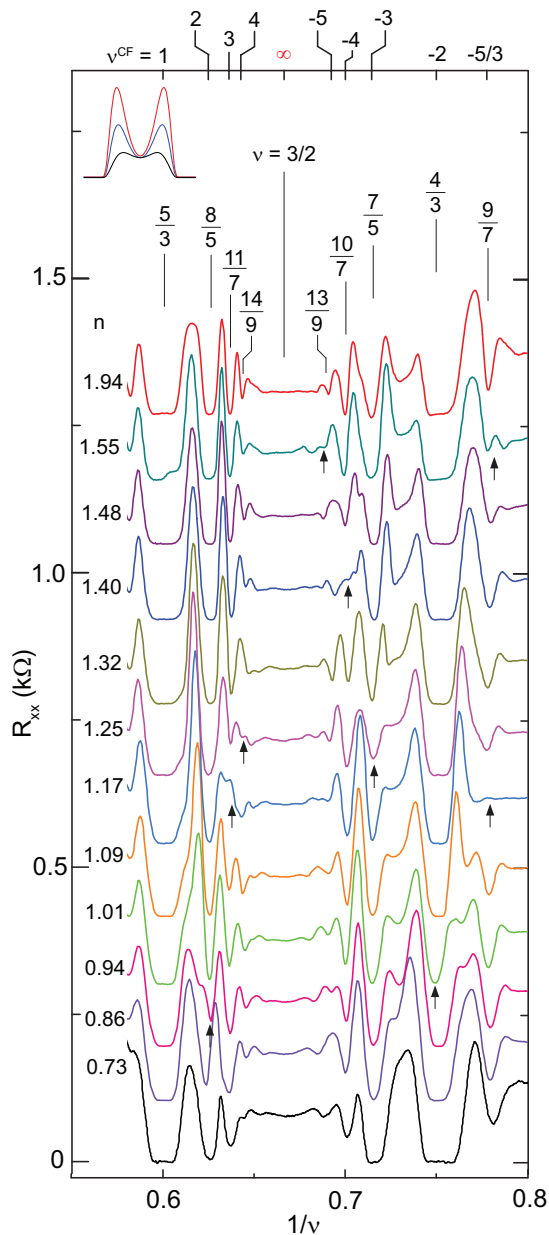}
\caption{(color online) $R_{xx}$ traces for 2D electrons confined to
  a 65-nm-wide, GaAs QW near $\nu=3/2$. The density for each trace is
  indicated (in units of $10^{11}$ cm${^{-2}}$), and traces are
  shifted vertically for clarity. Top inset shows the ($B=0$)
  calculated charge distributions at $n=0.73$, 1.40 and 1.94$\times
  10^{11}$ cm${^{-2}}$. Spin transitions of FQHSs, seen as a weakening
  or disappearing of $R_{xx}$ minima, are marked with arrows.}
\end{figure}

We made measurements on 2DESs confined to 31- to 65-nm-wide GaAs QWs
flanked by undoped AlGaAs spacer layers and Si $\delta$-doped layers,
which were grown by molecular beam epitaxy. The 2DESs have densities
($n$) ranging from 3.5 to 0.34, in units of $\times10^{11}$ cm$^{-2}$
which we use throughout this report, and very high mobilities, $\mu
\simeq$ 1,000 to 250 m$^2$/Vs. Each sample is a $4\times 4$ mm$^2$
cleaved piece, with alloyed InSn contacts at its four corners. We fit
the samples with an In back-gate and Ti/Au front-gate. By carefully
applying the front- and back-gate voltages, we can change $n$ while
keeping the QW symmetric. The measurements were carried out at
temperature $T\simeq$ 30 mK, and using low-frequency ($<40$ Hz)
lock-in techniques. We injected a very low measurement current $\sim
10$ nA to avoid polarizing the nuclear spins which might introduce an
effective nuclear field to our 2DES thus affecting the electron Zeeman
splitting \cite{Smet.PRL.2001, Hashimoto.PRL.2002, Kraus.PRL.2002,
  Tracy.PRL.2007}. We also checked the up and down magnetic field
sweeps to ensure the absence of any noticeable hysteresis.

In Fig. 1, we present longitudinal magnetoresistance ($R_{xx}$) traces
for a symmetric 65-nm-QW at different densities ranging from $n=0.73$
to 1.94. In our GaAs samples, $E_{\rm Z} \simeq 0.3 B\text{[T]}$ K
while $V_{\rm C} \simeq 50\sqrt{B\text{[T]}}$ K. Therefore, the ratio
$E_{\rm Z}/V_{\rm C}$ increases as $\sqrt{B}$ or $\sqrt{n}$ at a fixed
$\nu$. At a critical density $n_C$, when $E_{\rm Z}/V_{\rm
  C}=\alpha_{\rm C}$, a FQHS at a given $\nu$ makes a transition from
partial to full spin-polarization, signaled by a weakening or
disappearance of the $R_{xx}$ minimum and its reappearance at larger
$n$. We mark such transitions of FQHSs in Fig. 1 with arrows. The
$\nu=4/3$ FQHS ($\nu^{\rm CF}=-2$), e.g., is strong at the lowest
densities but becomes weak at $n=1.01$ and strong again at higher
$n$. We then identify 1.01 as $n_C$ for the $\nu=4/3$ FQHS ($\nu^{\rm
  CF}=-2$). At somewhat higher $n=1.25$ the $\nu=7/5$ FQHS ($\nu^{\rm
  CF}=-3$) exhibits a similar transition, followed by one for the
$\nu=10/7$ state ($\nu^{\rm CF}=-4$) at yet a higher $n=1.40$. On the
other side of $\nu=3/2$, we observe similar transitions for the FQHSs
at $\nu=8/5$, 11/7, and 14/9 ($\nu^{\rm CF}=2, 3, 4$) at $n=$ 0.94,
1.17, and 1.25, respectively.

In Fig. 2 we summarize the measured $n_C$ in the 65-nm-QW for each
FQHS as it becomes fully spin-polarized (closed square symbols). The
black dotted line represents the phase boundary above which all the
FQHSs in this sample are fully spin-polarized. It is clear in Fig. 2
that $n_C$ for this QW increases as $|1/\nu^{\rm CF}|$ decreases,
resulting a ''tent''-like shape for the phase boundary, with a maximum
at $\nu=3/2$ ($\nu^{\rm CF}=\infty$). This behavior has been observed
previously for the spin- \cite{Du.PRL.1995, Kukushkin.PRL.1999} or
valley-polarization \cite{Padmanabhan.PRB.2009} of the FQHSs, and is
also predicted theoretically \cite{Park.PRL.1998}. Below this
boundary, FQHSs can show several transitions as the CFs become
progressively more polarized with increasing $E_{\rm Z}/ E_C$
\cite{Du.PRL.1995}. An example of such multiple transitions is seen in
Fig. 1, where we identify two transitions for the $\nu=9/7$ FQHS
($\nu^{\rm CF}=5/3$) at densities $n=1.17$ and 1.55, the latter
corresponding to $n_C$ for the transition to a fully spin-polarized
state. Note in Fig. 2 plot that this $n_C$ is higher than $n_C$ for
the $\nu=4/3$ FQHS, suggesting that a second ''tent'' develops around
the even-denominator filling $\nu=5/4$. The FQHSs seen at fractional
$\nu^{\rm CF}$ correspond to higher-order CFs, and we will discuss
their spin transitions elsewhere.

\begin{figure}[htbp]
\includegraphics[width=.43\textwidth]{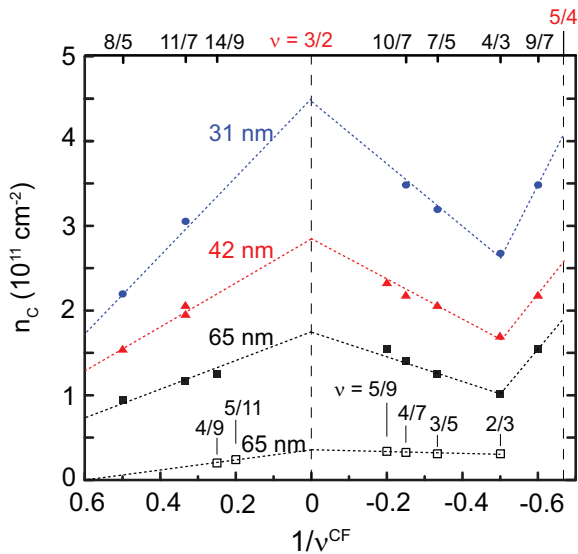}
\caption{(color online) Summary of the measured $n_C$ above which the
  FQHSs become fully spin-polarized. The dotted lines, drawn as
  guides to the eye, represent the phase boundaries that separate the
  partially-polarized (below) from the fully-polarized states
  (above). Data are shown for FQHSs near $\nu=3/2$ (closed symbols),
  and near $\nu=1/2$ (open symbols). The error bar for $n_C$ is about
  $\pm 5\%$ for FQHSs near $\nu=3/2$ and $\pm 10\%$ for those near
  $\nu=1/2$.}
\end{figure}

We performed similar experiments on the 31- and 42-nm-QWs, and
summarize the measured $n_C$ in Fig. 2. Clearly, $n_C$ strongly
depends on the QW well-width. For example, the $\nu=7/5$ state becomes
fully spin-polarized when $n\gtrsim 1.25$ in the 65-nm-QW, while it
remains partially polarized until $n$ reaches $\simeq$ 2.05 in the
42-nm-QW or $\simeq$ 3.19 in the 31-nm-QW. For the 42-nm-QW, we
measured $n_C$ in two different wafers with very different as-grown
densities, 1.8 and 2.9, whose densities can be tuned from 1.4 to 2.0
and 2.0 to 3.0, respectively. The fact that both samples have very
similar $n_C$ at $\nu=11/7$ confirms that, for a symmetric GaAs QW
with a given well-width, $n_C$ for a particular FQHS spin-polarization
transition is indeed an intrinsic property of the 2DES.

Figure 3 summarizes $\alpha_{\rm C}$ deduced from Fig. 2 data. In
Fig. 3, we also include $\alpha_{\rm C}$ reported for a GaAs/AlGaAs
heterojunction sample \cite{Du.PRL.1995}. The heterojunction sample
has a very small layer-thickness, typically $\simeq$ 0.1 $l_B$, and is
closer to an ideal, zero-thickness 2DES. It has a larger $\alpha_{\rm
  C}$ and, in Ref. \cite{Du.PRL.1995}, an additional parallel magnetic
was applied to enhance $E_{\rm Z}$ and reach the transition to full
spin-polarization. In Fig. 3 it is clear that $\alpha_{\rm C}$
decreases significantly as the 2DES layer-thickness increases. As we
discuss below, this strong thickness dependence of the phase boundary
stems from the softening of the Coulomb interaction when the electron
layer thickness becomes comparable to or larger than $l_B$.

An accurate assessment of the finite-layer-thickness effect requires
taking the shape of charge distribution into account. For a
semi-quantitative discussion we use the simple parameter
$\lambda/l_B$, where $\lambda$ is the standard deviation of the
electron's transverse position, as a measure of the thickness. To
determine $\lambda$, we performed calculations of the charge
distribution (at $B=0$) by solving the Schroedinger and Poisson
equations self-consistently and show examples of the resulting charge
distributions above Fig. 4. Note that when the QW width is large, the
2DES has a bilayer-like charge distribution at high densities but in
all cases $\lambda$ has a well-defined value. In Fig. 4, we plot our
measured $\alpha_{\rm C}$ as a function of $\lambda/l_B$ for the
$\nu=7/5$ FQHS. For the heterojunction sample, which has a very thin
layer-thickness, $\lambda/l_B\simeq 0.1$, while in our QW samples,
$\lambda$ is comparable to $l_B$. Figure 4 reveals that $\alpha_{\rm
  C}$ for the 7/5 FQHS monotonically decreases from about 0.03 in the
heterojunction sample ($\lambda/l_B\simeq 0.1$) to about 0.012 in the
65-nm-wide QW ($\lambda/l_B\simeq 1.1$). The same trend is also seen
for the other FQHSs.

\begin{figure}[htbp]
\includegraphics[width=.43\textwidth]{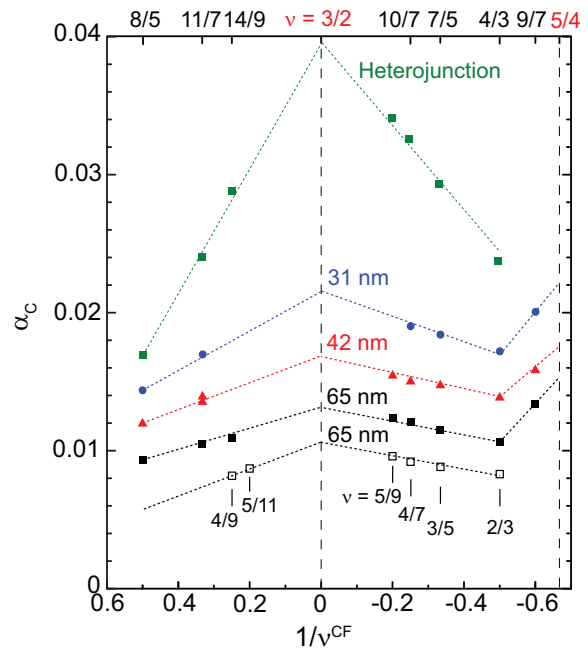}
\caption{(color online) Summary of $\alpha_{\rm C}$, the critical
  ratio of the Zeeman to Coulomb energies, needed to fully
  spin-polarize FQHSs at different $\nu$ and in different
  samples. Dotted lines represent the phase boundaries that separate
  the partially- (below) and fully-polarized (above) states. Closed
  (open) symbols are data measured near $\nu=3/2$ ($\nu=1/2$).}
\end{figure}

\begin{figure}[htbp]
\includegraphics[width=.45\textwidth]{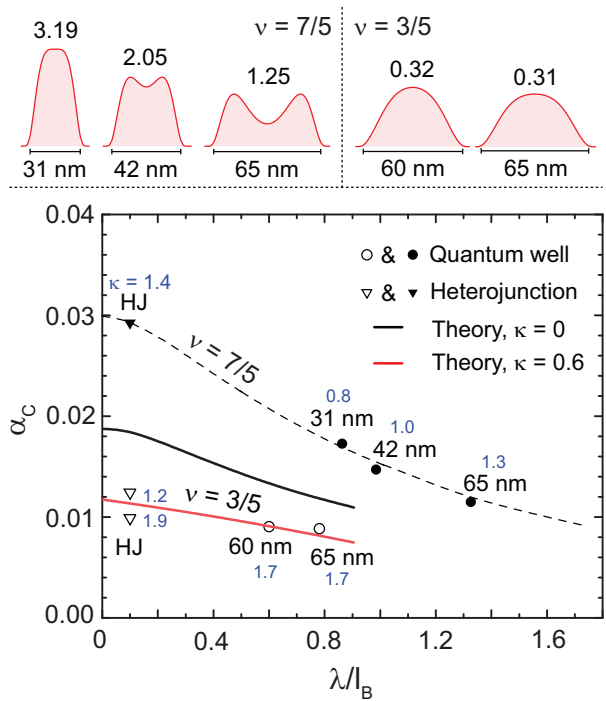}%
\caption{(color online) Measured critical polarization energy
  $\alpha_{\rm C}$ as a function of the layer-thickness, parameterized
  by $\lambda/l_B$, at filling factors $\nu = 7/5$ (solid symbols) and
  its particle-hole counterpart at $\nu = 3/5$ (open
  symbols). $\lambda$ is the standard deviation of the electron
  position, deduced from self-consistent charge distribution
  calculations, examples of which are shown in the panel above the
  figure. The dashed curve is a guide to the eye. The solid curves
  represent the theoretically calculated $\alpha_{\rm C}$ when
  $\kappa=0$ (black) and 0.6 (red). In this plot, we also include the
  $\kappa$ value (in blue) of experimental points.}
\end{figure}

In Figs. 2 and 3, we include the measured $n_C$ and $\alpha_{\rm C}$
for several FQHSs near $\nu=1/2$, namely those at $\nu=2/3$, 3/5, 4/7,
5/9, 4/9 and 5/11. The data were taken in another 65-nm-QW with a very
low as-grown density of 0.34. The charge distribution in this
low-density sample is single-layer-like and thinner than the higher
density 65-nm-QW sample (see Fig. 4 top panel), so that there is less
softening of the Coulomb interaction. However, instead of having a
larger $\alpha_{\rm C}$ compared to their particle-hole counterparts
near $\nu=3/2$, the FQHSs near $\nu=1/2$ have about 20\%
\textit{smaller} $\alpha_{\rm C}$ (see Fig. 3). In a more quantitative
comparison, when we normalize $\lambda$ to $l_B$, this discrepancy
becomes even larger. This mismatch is seen vividly in Fig. 4 where we
plot $\alpha_{\rm C}$ for the $\nu=3/5$ FQHS measured in
heterojunction samples \cite{Engel.PRB.1992, Kukushkin.PRL.1999}, and
in two QWs with $W=60$ and 65 nm. Since the $\nu=3/5$ and 7/5 FQHSs
both correspond to $\nu^{\rm CF}=-3$, they are expected to have the
same $\alpha_{\rm C}$ if particle-hole symmetry holds. However, as
seen in Fig. 4, $\alpha_{\rm C}$ for $\nu=3/5$ is less than half of
$\alpha_{\rm C}$ for $\nu=7/5$ at the same $\lambda/l_B$, implying
that the particle-hole symmetry is broken. We add that a very similar
discrepancy was recently observed for the \textit{valley}-polarization
energies of the FQHSs at $\nu=2/3$ and 4/3 in 2DESs confined to AlAs
QWs \cite{Padmanabhan.PRB.2010}.

Next we present results of our theoretical calculations. To include
the effect of finite $W$, we took $\cos^2(\pi z/W)$ as the shape of
the density profile in the transverse direction, for which
$\lambda\simeq 0.18~W$. Note that this simple model disregards the
double-humped density profile for large $W$ and $n$ (see Fig. 4 top
panel). We obtain the energies of the fully and partially spin
polarized states at different $\nu$ by extrapolating the finite system
results to the thermodynamic limit; the resulting $\alpha_{\rm C}$ is
shown in Fig. 4 (solid black curve) for $\nu=3/5$ or 7/5. It
reproduces the overall trend of decreasing $\alpha_{\rm C}$ with
increasing $\lambda/l_B$, as expected from a softening of the Coulomb
interaction due to the finite width. A quantitative discrepancy
remains, however, the sign of which depends on $\nu$.

We then include the effect of LL mixing. LL mixing modifies the
two-body interaction, while also producing an effective three-body
interaction (in a model that projects into the lowest LL); the latter
is more relevant here as it breaks the particle-hole symmetry. The
two- and three-body pseudopotentials have been obtained in a number of
articles in a perturbative scheme in the parameter $\kappa=V_{\rm
  C}/\hbar \omega_c$, where $\hbar\omega_c$ is the cyclotron energy
separation between LLs \cite{Bishara.PRB.2009, Simon.PRB.2013,
  Sodemann.PRB.2013, Note2}. We use the pseudopotentials given in
Ref. \footnote{Michael R. Peterson and Chetan Nayak, Phys. Rev. B 87,
  245129 (2013). We use $V^{(3)}_{1/2,3}= 0.0423$ for $\lambda=0$
  (M.R. Peterson, private communication) and assume that its magnitude
  scales with $\lambda$ in the same fashion as $V^{(3)}_{3/2,3}$.},
which are obtained perturbatively in $\kappa$ for finite widths
assuming a form of $\cos(\pi z/w)$ for the transverse wave function.
We keep all three-body pseudopotentials $V^{(3)}_{S,m}$ up to relative
angular momentum $m=3$ for spins $S=1/2$ and 3/2 \cite{Note2}. To
obtain the new $\alpha_{\rm C}$, we perform exact diagonalization
including these short range ($m\leq 3$) triplet pseudopotentials for
systems with $N=5$, 8, 11 (9, 12, 15, 18) for partially (fully) spin
polarized 3/5, and $N=13$, 20, 27 (25, 32, 39, 46) for the
corresponding 7/5 FQHSs. At 3/5, the correction to $\alpha_{\rm C}$ is
linear in $\kappa$ for up to $\kappa\approx 1$. Furthermore, LL mixing
\emph{depresses} $\alpha_{\rm C}$. The calculated $\alpha_{\rm C}$ for
$\kappa=0.6$, shown in Fig. 4 for $\nu=3/5$ as a red curve, show good
agreement with the experimental data. But note that $\kappa\gtrsim 1$
for the data, suggesting that the experiments are beyond the region
where a perturbative treatment is applicable, and the correction
presumably saturates with increasing $\kappa$.

The theoretical behavior is less clear for $\nu=7/5$.  Here, the
correction in $\alpha_{\rm C}$ is not proportional to $\kappa$ even
for $\kappa\approx 0.5$, implying that the ground-state itself is
affected significantly by LL mixing for this value of
$\kappa$. Furthermore, an irregular size dependence makes the
extrapolation to the thermodynamic limit unreliable, suggesting the
need for larger systems and perhaps also for pseudopotentials beyond
$m=3$. As a result, we are not able to ascertain reliably the
correction to $\alpha_{\rm C}$ for 7/5 that can be compared
semi-quantitatively to experiments; nonetheless, our calculations
clearly demonstrate that LL mixing causes at 7/5 an \emph{increase} in
$\alpha_{\rm C}$, in agreement with the experimental observation. We
studied other states near $\nu= 1/2$ (2/5, 3/7, 2/3) and $\nu=3/2$
(8/5, 11/7, 4/3) and found a behavior similar to 3/5 and 7/5. Our
results imply that LL mixing causes a significant renormalization of
the polarization mass of CFs (for a fixed $\lambda/l_B$), increasing
it in the vicinity of $\nu=1/2$ but lowering it near $\nu=3/2$.

In summary, our systematic study of 2DESs confined to symmetric GaAs
QWs reveal that the critical Zeeman energies where FQHSs become fully
spin polarized depend substantially on finite layer thickness and,
more importantly, on LL mixing, which breaks particle-hole
symmetry. Our results thus provide fundamental insight into the nature
of the three-body interaction terms induced by LL mixing.

\begin{acknowledgments}
  We acknowledge support from the DOE BES (DE-FG02-00-ER45841) for
  measurements, and the Gordon and Betty Moore Foundation (GBMF2719),
  the Keck Foundation, and the NSF (DMR-1305691 and MRSEC DMR-0819860)
  for sample fabrication. The experiments were partly performed at the
  National High Magnetic Field Laboratory, which is supported by NSF
  Cooperative Agreement No. DMR-1157490, by the State of Florida, and
  by the DOE. For our theoretical work, we acknowledge the DOE grant
  no. DE-SC0005042 for J. K. Jain, and the Polish NCN grant
  2011/01/B/ST3/04504 and EU Marie Curie Grant PCIG09-GA-2011-294186
  for A. W\'ojs. The computations were performed using computing
  facilities of the Cyfronet and WCSS, both parts of PL-Grid
  Infrastructure. We thank S. Hannahs, E. Palm, J. H. Park,
  T. P. Murphy, and G. E. Jones for technical assistance, and
  M. R. Peterson for sharing with us his unpublished results.
\end{acknowledgments}

\bibliography{../bib_full}
\end{document}